\documentclass{article}

\usepackage{arxiv}

\usepackage[utf8]{inputenc} 
\usepackage[T1]{fontenc}    
\usepackage{hyperref}       
\usepackage{url}            
\usepackage{booktabs}       
\usepackage{amsfonts}       
\usepackage{nicefrac}       
\usepackage{microtype}      
\usepackage{lipsum}
\usepackage{graphicx}
\usepackage{amsmath}

\title{Mean Gradient Descent:\\ An optimization approach for single-shot interferogram analysis}

\author{
  Sunaina \\
  Department of Physics\\
  Indian Institute of Technology Delhi\\
  New Delhi 110016 India \\
   \And
 Mansi Butola \\
 Department of Physics\\
 Indian Institute of Technology Delhi\\
 New Delhi 110016 India \\
 \AND
 Kedar Khare \\
 Department of Physics\\
 Indian Institute of Technology Delhi\\
 New Delhi 110016 India \\
 \texttt{kedark@physics.iitd.ac.in} \\
}

\begin{document}
\maketitle

\begin{abstract}
Complex object wave recovery from single-shot interference pattern is an important practical problem in interferometry and digital holography. The most popular single-shot interferogram analysis method involves Fourier filtering of cross-term but this method suffers from poor resolution. For obtaining full pixel resolution, it is necessary to model the object wave recovery as an optimization problem. The optimization approach typically involves minimizing a cost function consisting of a data consistency term and one or more constraint terms. Despite its potential performance advantages, this method is not used widely due to several tedious and difficult tasks such as empirical tuning of free parameters. We introduce a new optimization approach ``Mean gradient descent (MGD)'' for single-shot interferogram analysis that is simple to implement, robust and does not require any free parameters. The MGD iteration does not try to achieve minimization of any cost function but instead aims to reach a solution point where the data consistency and the constraint terms balance each other. This is achieved by iteratively progressing the solution in the direction that bisects the descent directions associated with the error and constraint terms. Numerical illustrations are shown for recovery of a step phase object from its corresponding off-axis as well as on-axis interferograms simulated with multiple noise levels. Our results show full pixel resolution as evident from the recovery of phase step and excellent rms phase accuracy relative to the ground truth phase map. The concept of MGD as presented here can potentially find applications to wider class of optimization problems. 
\end{abstract}


\section{Introduction}
Demodulation of interference fringes is of importance to a wide ranging applications in optics \cite{Hariharan} including optical metrology, digital holography for live cell imaging, surface inspection, particle-field holography, astronomical imaging etc. Currently the interference fringes in all these applications are predominantly recorded on CCD or CMOS array sensors that are readily available. The digitally recorded fringe pattern is then numerically processed for recovering the complex-valued object wave of interest. Traditionally there are two main methods that are used for interferogram analysis. For off-axis interference pattern the dc and the cross-terms in the interference pattern can be nominally separated in Fourier space and thus the cross term can be obtained by Fourier space filtering \cite{Takeda, Kreis}. This method requires single interference pattern but as explained later, the Fourier space filtering operation inherently causes loss of resolution. Phase shifting interferometry \cite{Creath,Yamaguchi} on the other hand can achieve full pixel resolution using a multi-shot interferogram recording approach and requires stringent vibration isolation. The CCD/CMOS arrays available today have good sensitivity so that just a milli-second of exposure (during which the recording is not affected much by ambient vibrations) is often sufficient to record good quality interferograms with nominal tabletop interferometer setups. In view of this it is highly desirable to have a practical algorithm that can demodulate a single-shot interferogram without compromising on resolution and accuracy.

In this work we are primarily interested in complex-valued object recovery from a single-shot image plane digital hologram/interferogram represented as:
\begin{equation}\label{hologram}
    H = |R|^2 + |O|^2 + RO^{*} + R^{*}O,
\end{equation}
Here the hologram $H$, the reference beam $R = |R|e^{i\phi_R}$ and the object beam $O = |O|e^{i\phi_O}$ are two-dimensional functions of the pixel coordinates $(x,y)$ at the sensor plane. Further we will assume that the reference beam $R$ (both its magnitude $|R|$ and phase $\phi_R$) is known and the problem is to recover the object wave function $O$ from the single-shot hologram. We observe that since we are trying to fit two functions $|O(x,y)|$ and $\phi_O (x,y)$ to a single hologram frame $H(x,y)$, the problem does not have a unique solution and additional constraint is required in order to obtain a desirable solution. For example in the present case of image plane holography, it is expected that the object of interest to be imaged will typically have sharp edges and hence constraints like minimal Total Variation (TV) may be applied to the object field $O(x,y)$. Such constraints can be included effectively if the object wave recovery is modeled as an optimization problem.  

The first optimization framework for interferogram analysis may be considered to be the regularized phase tracking (RPT) method\cite{Servin, Servin2014}. In this approach the amplitude and phase of an unknown object are separately recovered pixel by pixel by fitting a local polynomial model for the unknown object wave after the removal of low frequency intensity terms from the interferogram data. In another work on complex wave retrieval algorithm \cite{Liebling} the phase and amplitude of the object are separately recovered by local least-square fitting with a variable sized window after changing the non-linear variable equation of hologram to a set of linear equations. A nominal constrained optimization approach using the complex (or Wirtinger) derivatives \cite{Brandwood} for functional gradients was proposed in \cite{Kedar}. This approach demonstrated recovery of object wave from a single-shot off-axis hologram even when the dc and the cross terms showed significant overlap in Fourier space. This method was further used to demonstrate highly accurate phase recovery from low light level interference data \cite{Mandeep2015}. An alternating amplitude and phase optimization strategy was developed in \cite{Bouman} which is an interesting approach that again demonstrated resolution improvement over the Fourier transform method. The optimization methodology in \cite{Kedar, Mandeep2015} was modified further in \cite{Mandeep2017} where an adaptive optimization approach was proposed. In the optimization framework the problem of reconstruction of the object wave $O$ typically modelled as minimization of a two-term cost function:
\begin{align}\label{Cost}
C(O,O^*) &= ||H - |O+R|^2||^{2}_{2} + \alpha \: TV(O,O^*) \\
        &= C_{err}(O,O^*) + \alpha \: C_{TV}(O,O^*),
\end{align}
Here $C, C_{err}, C_{TV}$ are functions of the hologram $H$, the reference beam $R$ and the unknown complex object wave $O$. The first term $C_{err}$  refers to the L2-norm squared error or data-consistency term and $C_{TV}$  refers to the TV penalty. The positive number $\alpha$ determines the weight between two terms of the cost function. The definition of TV we use for the present work is\cite{Shi}:
\begin{equation}
    TV(O,O^*) = \sum_{i= all pixels} [\;\; |\nabla_x O_i| \: + |\nabla_y O_i| \;\;] \:,   
\end{equation}
where, $\nabla_x$ and $\nabla_y$ are the $x$ and $y$ partial derivative operators respectively. In \cite{Mandeep2017}, it was shown that starting with any initial guess $O^{(0)}$ for the solution, if the cost function is iteratively reduced via a gradient descent scheme, the quality of the resultant solution depends on the numerical value of $\alpha$. The parameter $\alpha$ therefore needs to be determined empirically which is often not desirable. 

In the present work we propose a novel approach - Mean Gradient Descent (MGD) - that does not require any weight parameter $\alpha$. The aim of MGD is not to achieve minimum of any cost function as in Eq. (\ref{Cost}) but to instead achieve balance between the two terms $C_{err}$ and $C_{TV}$. Our methodology is inspired by a successful image reconstruction algorithm ASD-POCS \cite{Pan} for X-ray computed tomography. In an earlier work \cite{Mandeep2017}, a methodology very similar to ASD-POCS  was employed for the single-shot interferogram analysis. The main idea in ASD-POCS is to reach a solution point where the directions $- \nabla_{O^{\ast}} C_{err}$ and $- \nabla_{O^{\ast}} C_{TV}$ corresponding to descent directions for the reduction in $C_{err}$ and $C_{TV}$ make an obtuse angle. The descent directions corresponding to the two terms of the cost function can thus be thought to achieve an equilibrium. ASD-POCS achieves this balance by alternating minimization of $C_{err}$ and $C_{TV}$. Starting with $O = O^{(n)}$ the error term $C_{err}$ is reduced leading to an intermediate solution $O^{(n)}_{int}$. The $C_{TV}$ associated with this intermediate solution is then recursively reduced to obtain the next guess $O^{(n+1)}$ such that the distances $d_1 = ||O^{(n)} - O^{(n)}_{int}||_2$  and $d_2 = ||O^{(n+1)} - O^{(n)}_{int}||_2$ are approximately matched in every iteration. Once the solution reaches near the optimal point, ensuring that $d_1 \approx d_2$ in further iterations causes negligible change in the solution as the two descent directions oppose each other. 

We show here that reaching the balance point as in ASD-POCS is possible without employing an alternating minimization scheme but by iteratively progressing the solution in a direction that bisects that of the two functional gradients. The MGD iteration is computationally much simpler as compared to the alternating minimization approach. While MGD is discussed here in the context of single-shot interferogram analysis, we believe that it may be useful to two term optimization problems in general. In the context of interferogram analysis we show that procedure is uniformly applicable to various interferometric configurations (off-axis as well as on-axis) and thus amenable to be used with multiple digital holography/ interferometry system configurations.

The paper is organized as follows. In section 2 we provide a detailed description of MGD algorithm with the help of numerical illustrations for different noise levels present in the interferogram data corresponding to off-axis plane reference beam and step phase object. In section 3, we demonstrate the performance of MGD iteration for on-axis and near on-axis spherical reference beam configurations of the interferogram setup. Finally,in section 4, we summarize our observations and provide our insights over the results obtained by MGD algorithm.

\section{Description of MGD iteration}
For simplicity we describe the MGD iteration with an illustration of single-shot off-axis interferogram.  For object wave in the interferogram plane, we use a square-shaped binary phase object on a $500 \times 500$ pixel grid. We have taken unit amplitude across the entire object and a step phase of $2\pi/3 $ radians inside the square area of $250 \times 250$ as shown in Fig. 1(a). A step phase object is used here as we are interested in image plane holography where the object wave $O$ may contain sharp edges. Although in a realistic case the edge sharpness will be limited by numerical aperture of the imaging system, in this case we have assumed a phase object with ideal phase step without any band-limit which is generally considered to be a hard problem for traditional Fourier filtering approach. In the methods based on optimization approach the step phase reconstruction is still a hard problem due to involvement of empirical parameters \cite{Legarda, Galvan}. The reference beam is an off-axis plane wave given by $R = \exp[i 2\pi (f_1 x + f_2 y)$ with $f_1 = f_2 = 0.04$ /pixel. The corresponding interferogram is shown in Fig. 1(b). The interferogram has been simulated with Poisson noise equivalent to an average  light level $10^4$ photon/pixel. Since the dc and cross-term peaks of this interferogram are separated in Fourier space, the cross term may be filtered as shown in Fig. 1(c) where a filter of radius $0.5$ times the distance between the dc and cross terms peaks in Fourier space has been used for illustration. A Hamming window is also applied to the filtered cross term prior to computing the inverse Fourier transform in order to mitigate ringing artifacts. The resultant phase map is shown in Fig. 1 (d) has poor resolution compared to the step phase object in Fig. 1(a). Recently a Hilbert transform processing methodology \cite{Baek} has been demonstrated which provides superior resolution than what is obtained in Fig. 1(d), however, this procedure still requires a band-limited object wave.
\begin{figure}[htbp]
    \centering
    \includegraphics[width=0.47\textwidth]{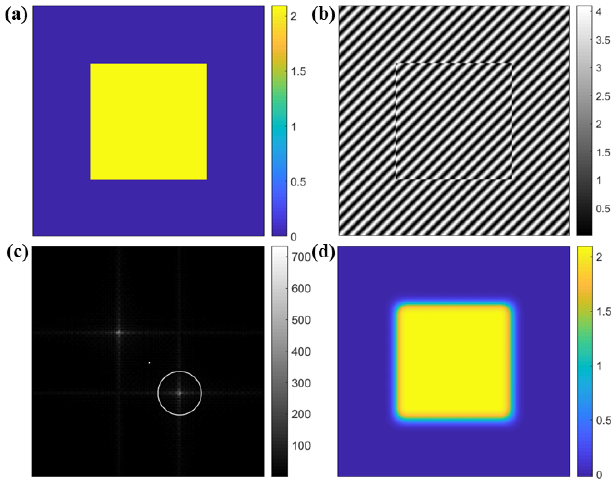}
    \caption{(a) Phase map of square shaped object with a step phase of $2\pi/3$ radians over $250 \times 250$ pixels defined on a $500 \times 500$ pixel grid. (b) Hologram of the object in (a) with a tilted plane reference beam simulated with Poisson noise for an average light level of $10^4$ photons/pixel. (c) Zoomed-in version of Fourier magnitude of hologram showing the circular filter of radius $0.5$ times the distance between dc and cross term peaks. For display purpose the Fourier magnitude of hologram is shown as $|H'(f_x,f_y)|^{0.5}$ (d) Phase image reconstructed by Fourier Transform method (FTM). }
    \label{fig:my_label}
\end{figure}

We will now proceed with a description of MGD iteration for recovering the complex object wave $O$ from a single-shot interferogram $H$. In general the goal of MGD is to find a solution $O$ such that costs associated with both data consistency $C_{err}$ and the Total Variation $C_{TV}$ as in Eq. (\ref{Cost}) simultaneously achieve minimal numerical values and that additionally the two functional gradients associated with these costs balance each other. For a given guess solution $O$, we start by defining two unit vectors:
\begin{equation}\label{erred_uv}
    \hat{\bf u}_1 =\frac{\nabla_{O^*} C_{err}(O,O^*)}{|| \nabla_{O^*}C_{err}(O,O^*)||_2}
\end{equation}
and 
\begin{equation}\label{tvred_uv}
    \hat{\bf u}_2 =\frac{\nabla_{O^*} C_{TV}(O,O^*)}{|| \nabla_{O^*}C_{TV}(O,O^*)||_2}.
\end{equation}
Here, the two functional gradients (or Wirtinger derivatives) are defined as:
\begin{equation}
    \nabla_{O^*} C_{err}(O,O^*) = -2(H-|O + R|^2)(O+R),
\end{equation}
and
\begin{equation}
    \nabla_{O^*} C_{TV}(O,O^*) = - \nabla.[\frac{\nabla_x O}{|\nabla_x O|}\hat{x}+\frac{\nabla_y O}{|\nabla_y O|}\hat{y}].
\end{equation}
Next we introduce a vector ${\bf u}$ that is along the mean direction that bisects $\hat{\bf u}_1$ and $\hat{\bf u}_2$:
\begin{equation}\label{mean_dir}
    {\bf u} = \frac{\hat{\bf u}_1\:+ \hat{\bf u}_2}{2},
\end{equation}
and consider an iteration of the form:
\begin{equation}\label{mgd}
        O^{(n+1)} = O^{(n)} \: - t\:||O^{(n)}||_2\: [{\bf u}]_{O=O^{(n)}}. 
\end{equation}
Here the parameter $t$ denotes the step size in units of the norm $||O^{(n)}||_2$ of the guess solution after $n$ iterations. Note that since $\hat{\bf u}_1$ and $\hat{\bf u}_2$ are unit vectors, for any arbitrary value of $t$, the changes in the solution 
\begin{align}
   D_{1,n} &= ||\;\;\frac{t}{2}\:||O^{(n)}||_2\: [\hat{\bf u}_1]_{O=O^{(n)}} \;\;||_2, \nonumber \\
   D_{2,n} &= ||\;\;\frac{t}{2}\:||O^{(n)}||_2\: [\hat{\bf u}_2]_{O=O^{(n)}} \;\;||_2
\end{align}
due to the progression in the error and TV reducing directions are always guaranteed to be equal. The iteration is much easier computationally as compared to the alternating minimization scheme required for ASD-POCS type algorithms. The scheme for selecting $t$ will be explained later. In order to understand the progression of the solution by MGD, we examine the behaviour of $C_{err}$, $C_{TV}$ and the angle $\theta$ between the directions of $\hat{\bf u}_1$ and $\hat{\bf u}_2$ as the iterations progress. Since the two vectors $\hat{\bf u}_1$ and $\hat{\bf u}_2$ are complex-valued, for the purpose of calculating angle between them, we form new vectors by concatenating their real and imaginary parts:
\begin{align}\label{v1v2}
{\bf v}_1 = [\textrm{real}(\hat{\bf u}_{1j}) , \textrm{imag}(\hat{\bf u}_{1j})] \nonumber \\
{\bf v}_2 = [\textrm{real}(\hat{\bf u}_{2j}) , \textrm{imag}(\hat{\bf u}_{2j})].
\end{align}
Here the index $j$ runs over computational window size ($j = 1, 2, ..., (500)^2$). The angle between ${\bf u}_1$ and ${\bf u}_2$ is then defined as:
\begin{equation}\label{theta}
    \theta = \arccos[\frac{ {\bf v}_1 \cdot {\bf v}_2 } {||{\bf v}_1 ||_2 ||{\bf v}_2 ||_2}].
\end{equation}

As seen in illustrations below, we observe that following this scheme leads to successive solutions where
both $C_{err}$ and $C_{TV}$ nominally decrease and eventually, the angle between $\hat{\bf u}_1$ and $\hat{\bf u}_2$ becomes obtuse implying that the resultant solution balances the two terms $C_{err}$ and $C_{TV}$. We term this scheme as ``Mean Gradient Descent'' in view of the definition Eq. (\ref{mean_dir}) of vector ${\bf u}$ and that the two costs are seen to nominally progress to their minimal possible values. 
\begin{figure}[htbp]
    \centering
    \includegraphics[width=0.47\textwidth]{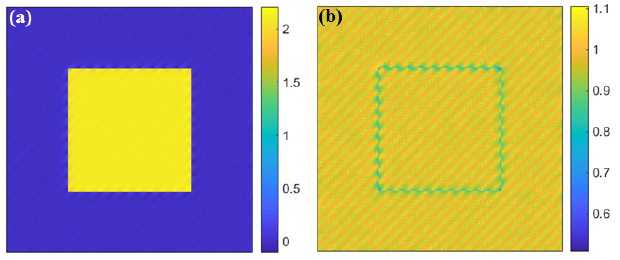}
    \caption{(a) Phase and (b) amplitude of solution for object wave after $200$ MGD iterations with step size $t$ kept fixed.}
\end{figure}
\begin{figure}[htbp]
    \centering
    \includegraphics[width=0.43\textwidth]{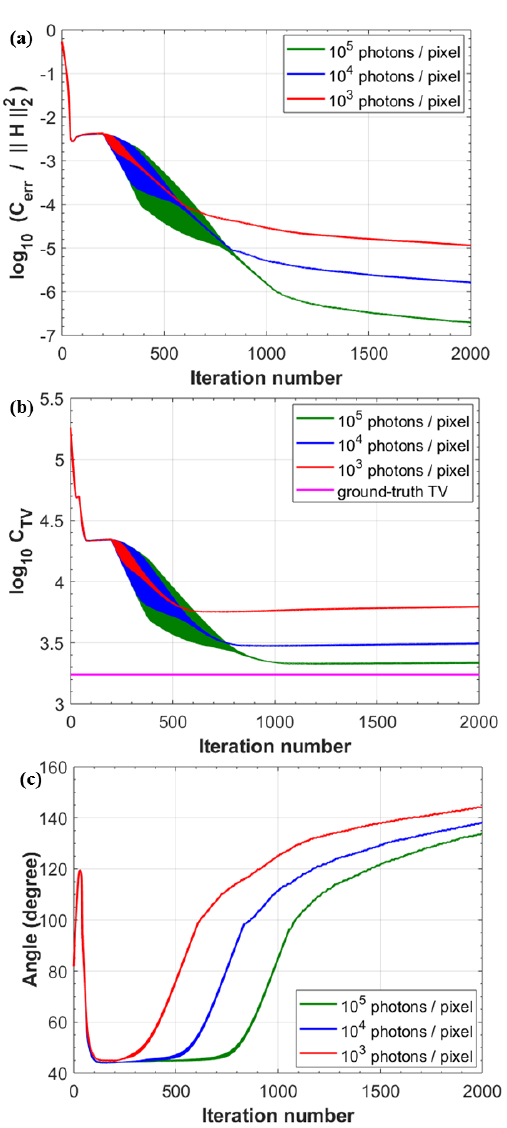}
    \caption{Behaviour of (a) logarithm of $C_{err}$ with iteration number, and (b) logarithm of $C_{TV}$ with iteration number for three light levels of $10^{3}$,$10^4$ and $10^5$ photons/pixel.(c) Variation of angle between $\hat{\bf u}_1$ and $\hat{\bf u}_2$ with number of iterations corresponding to the average light levels of $10^{3}$,$10^4$ and $10^5$ photons/pixel. }
\end{figure}
For a typical illustration with Poisson noise corresponding to the average light level of $10^4$ photons/pixel in the hologram data frame, we initiate the MGD iteration with a random complex valued function with amplitude and phase distributed uniformly in $[0,1]$ and $[0,2\pi]$ respectively. As per Eq. (\ref{mgd}), in each iteration the solution is simply pushed in the direction $-{\bf u} = -(\hat{\bf u}_1\:+ \hat{\bf u}_2)/{2}$. In the following discussion we provide our thought process for selecting the step size $t$ which is kept constant for initial iterations and then reduced slowly.

As already explained before, the problem of determining amplitude $|O(x,y)|$ and phase $\phi_O (x,y)$ from a single hologram data frame $H(x,y)$ does not have a unique solution even when the reference beam $R(x,y)$ is known exactly. There can be multiple combinations of the amplitude and phase functions that may satisfy the hologram data. Starting with any random guess solution, if we reduce $C_{err}$ alone by a gradient descent scheme, we observe that we reach a solution representing a local minimum of $C_{err}$ that does not simultaneously have a low value for $C_{TV}$. The progression in the mean direction  however leads to moving away from such undesirable solutions. We note that since $\hat{\bf u}_1$ and $\hat{\bf u}_2$ are unit vectors, the maximum magnitude of ${\bf u}$ is equal to $1$. We therefore initiate $t$ with a nominal trial value of $0.1$, suggesting that the solution can maximally change in norm by $10 \%$ in a single iteration. When this initial value of $t$ is held constant for the first few hundred iterations, we find that the solution reaches close to the desirable solution in the solution landscape. The phase and amplitude maps corresponding to the resultant solution after $200$ iterations with a fixed $t$ value are shown in Fig. 2 (a), (b) respectively. We note that the solution has the expected features of a step phase object but additionally contains undesirable fringe-like artifacts. The blue curves in Fig. 3(a), (b), (c) show the plots for $C_{err}$, $C_{TV}$ and $\theta$ as a function of iteration number respectively. Fig. 3 shows plots of these quantities for three different noise levels as we will discuss later. For now we will concentrate on the blue curves in these plots that correspond to the Poisson noise corresponding to the average light level of $10^4$ photons/pixel. At the end of $200$ iterations we observe that the three curves for $C_{err}$, $C_{TV}$ and $\theta$ have nearly flattened. (The blue, red and green curves in Fig. 3 representing different noise levels are nearly overlapping in this region). This behaviour suggests that the solution has essentially stagnated. While the solution appears to be close to what we want (with some additional artifacts) the fixed value of $t$ is too large at this point and the solution is unable to approach the desired minimum in $C_{err}$ or $C_{TV}$. 

In the further iterations, we check the value of error term $C_{err}$. If the numerical value of $C_{err}$ has increased compared to that in the previous iteration, the step size $t$ is decreased slowly by a constant factor $0.99$. The plots in Fig. 3 (a), (b), (c) start showing interesting behaviour at this point - the numerical values of $C_{err}$ and $C_{TV}$ are seen to start nominally decreasing while the angle $\theta$ starts increasing as we reduce $t$. The oscillations in these curves right after $t$ starts reducing represents the fact that $t$ is still too large than what is desired and is being slowly adjusted to a lower value. It is important to note that every iteration simply involves a fixed straightforward computation of $\hat{\bf u}_1$ and $\hat{\bf u}_2$ followed by progressing the solution in the mean direction, thus the computational cost per iteration is minimal. 

In Fig. 4 (a),(c) and Fig. 4 (b),(d) we show the amplitude and phase maps of the resultant solutions after $500$ and $2000$ MGD iterations for the data with Poisson noise corresponding to the average light level of $10^4$ photons/pixel. The computational time for a MATLAB implementation on a $3.5$ GHz processor and $16$ GB RAM was observed to be approximately $0.07$ seconds per iteration. The profile of the phase function after $2000$ iterations along the dotted line in Fig. 4 (d) is plotted in Fig. 4 (e) and shows excellent recovery of the sharp edge in phase function. 
\begin{figure}[htbp]
    \centering
    \includegraphics[width=0.45\textwidth]{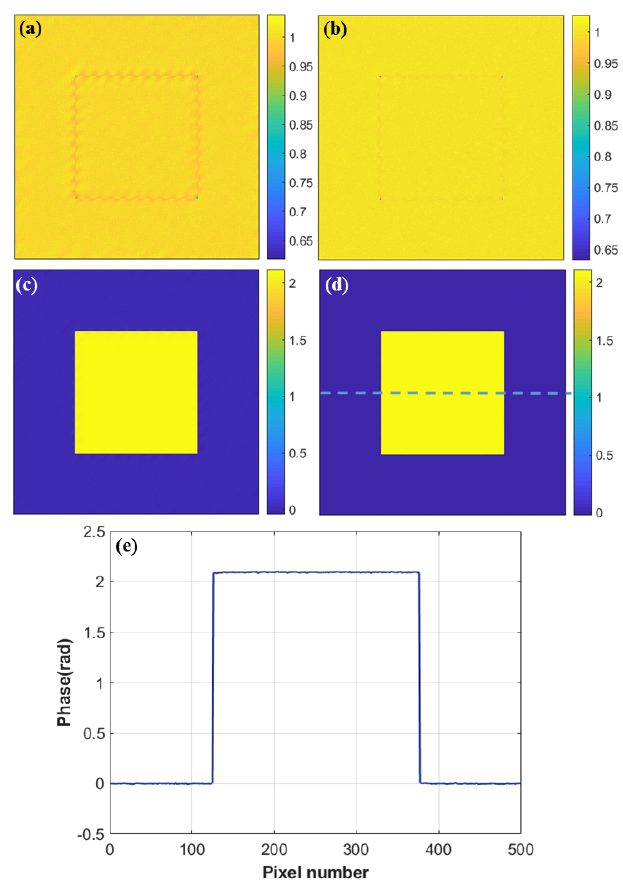}
    \caption{Progression of solution with MGD algorithm for Poisson noise realization with light level of $10^4$ photons/pixel. Amplitude of the solution after (a) $500$ and (b)$2000$ iterations. Phase reconstruction after (c)$500$ and (d)$2000$ iterations. (e) Phase profile of the resultant solution in (d) along the dotted line. Note that the solution contains sharp edges as compared to FTM solution shown in Fig. 1(d).  }
\end{figure}
In order to understand the sensitivity of MGD approach to noise we considered two additional Poisson noise realizations of the interferogram in Fig. 1(b) corresponding to an average light level of $10^3$ and $10^5$ photons/pixel. The behaviour of $C_{err}$, $C_{TV}$ and $\theta$ for these cases as a function of iterations is also shown in Fig. 3 (a), (b), (c) respectively (red and green curves). We observe that with increasing light level, the resultant numerical values of $C_{err}$ and $C_{TV}$ are consistently lower and the numerical value of $C_{TV}$ increasingly approaches the ground truth value of TV of the simulated step-phase object (shown in magenta). We note that the rise in angle $\theta$ after $200$ iterations when we start reducing step size $t$ is fastest for the data with the highest relative noise ($10^3$ photons/pixel). This is expected since the balancing of the two terms of the cost function should start happening at higher value of $C_{err}$ for data with higher noise.

Table 1 shows the RMS phase error between the ground truth phase map and the reconstructed phase map for the three noise realizations and the RMS error performance is best for the data with lowest relative noise as expected. The RMS error performance for all the three cases is excellent and in fact superior to the expected performance purely based on shot noise considerations\cite{Walls}. Superior performance compared to single-pixel based shot noise level is expected due to the sparsity of the object wave as already demonstrated in \cite{Mandeep2015}. A more detailed analysis of performance of MGD with respect to the light level and the sparsity of the object wave will be taken up in future. For random initial guess we observe that the initial direction between $\hat{{\bf u}}_1$ and $\hat{{\bf u}}_2$ was observed to be close to $90^\circ$ suggesting that the two directions were independent. As the iterations progressed, the angle $\theta$ was initially acute but eventually became obtuse and close to $180^\circ$ as the number of iterations was made very large. This behaviour of angle $\theta$ confirms our main motivation for using the MGD approach. 
\begin{table}[htbp]
\centering
\caption{\bf Phase rms error values after $2000$ iterations of MGD algorithm corresponding to three different noise levels added to the hologram data. }
\begin{tabular}{ |p{2cm}|p{1cm}|p{1cm}|p{1cm}|p{1cm}| p{1cm}| }
 \hline
Light level (N photons/pixel) & $10^3$ & $10^4$ & $10^5$ \\
 \hline
 RMS error (rad) &  $0.0125$ &$0.0042$  &$0.0021$ \\
\hline
Shot Noise level $1/\sqrt{N}$ (rad) & $0.0316$ & $0.010$ & $0.0032$ \\
\hline
\end{tabular}
\end{table}
\section{Performance of MGD for on-axis and near on-axis interferograms with spherical reference beam}
From the previous section it is clear that the working of MGD is independent of the various noise levels in the hologram data. It should also be noted that the MGD approach never utilized the fact that we were analyzing an off-axis interferogram. As long as the form of reference beam is known, MGD should be able to handle the hologram data in the same manner as the off-axis case as we illustrate in this section. We now test the evolution of MGD solution for two interferogram recording configurations shown in Fig. 5 (a), (b) where the reference beam is in the form of on-axis and near on-axis spherical beams. The on-axis spherical wave is taken of the form $R = \exp(i 2 \pi p (x^2 +y^2))$ and the near on-axis spherical wave has form $R = \exp(i 2 \pi q [(x-51)^2 +(y-51)^2])$ with $p = q = 0.0004/\textrm{pixel}^{2}$. The interferograms in Fig. 5(a), (b) are generated with Poisson random noise corresponding to average light level of $10^4$ photons/pixel. Note that for both on-axis and the near on-axis spherical reference beam configurations, the dc and the cross terms in the interferograms substantially overlap in the Fourier domain as clearly visible in Fig. 5(c),(d). As a result there is no effective Fourier filtering strategy (as in Fig. 1) that can separate out the object wave even in an approximate sense. The MGD iterations is independent of such considerations and high quality object wave reconstructions are obtained as shown in Fig. 5 (e), (f). The edge profiles corresponding to resultant solutions in  Fig. 5 (e), (f) are plotted in Fig. 5(g),(h) along the dotted lines respectively. The phase profiles clearly show the excellent step phase recovery for both the cases. The behaviour of $C_{err}$, $C_{TV}$ and angle $\theta$ for the two configurations as the iterations progress is observed to be similar to that of the previously illustrated off-axis case in Fig. 3(a)-(c) respectively. The RMS phase error, with respect to the ground truth solution, calculated after $2500$ MGD iterations for the reconstructed phase solutions in Fig. 5(e), (f) are $0.0040$rad and $0.0048$rad respectively which is similar to the numerical value in Table 1. The computational time per iteration for both the cases illustrated above is identical to the off-axis case as the steps involved in MGD are independent of the nature of interference pattern. From these results, MGD appears to be a robust methodology that uniformly works with multiple interferometric configurations and noise levels. The algorithm is therefore expected to have applicability for interferometric systems in various geometrical configurations. To the best of our knowledge, the MGD iteration as presented here has not been explored in the prior literature. At present we are reporting results of MGD with illustrative examples for lack of a formal proof for its convergence properties. Such proofs if possible will have to be worked out in future. Our numerical trials in this and the previous section however suggest that MGD works uniformly in a robust manner and provides excellent complex object wave recoveries from single-shot interferogram data. We believe that MGD as a concept may be useful beyond interferometry in a multitude of optimization problems that are similar in nature to the present problem as described in Eq. (\ref{Cost}).
\begin{figure}[htbp]
    \centering
     \includegraphics[width=0.45\textwidth]{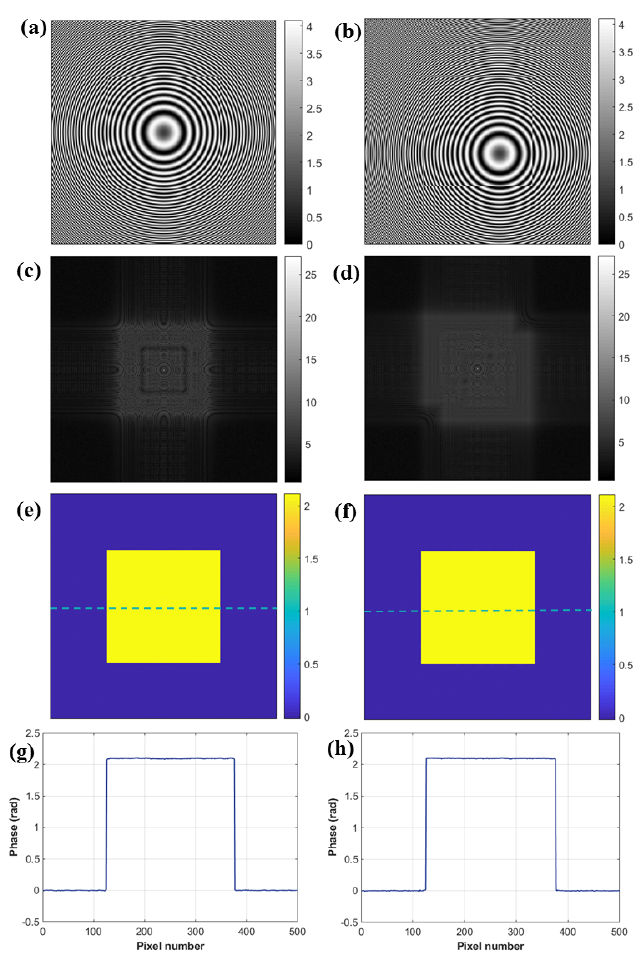}
    \caption{ Hologram of step phase object with (a) on-axis and (b) near on-axis spherical reference wave simulated with Poisson noise with an average light level of $10^4$ photons/pixel. (c),(d) Fourier magnitudes of the holograms in (a) and (b) respectively showing overlap between dc and cross terms. (e),(g) and (f),(h) are the reconstructed phase maps and their corresponding phase profiles along the dotted lines respectively after $2500$ MGD iterations.}
\end{figure}

\section{Discussion and future outlook}
In summary we have presented a new optimization approach that we call as Mean Gradient Descent (MGD) for single-shot interferogram analysis. Unlike the usual optimization approaches which aim to minimize a cost function, we aim to reach a solution point where the data consistency and constraint penalty functions balance each other. This is achieved by iteratively progressing the solution in a direction that bisects the descent directions for the data consistency (or error) and the penalty terms. This approach does not require any free parameters. MGD scheme works uniformly for varying noise levels in the data as well as data representing different interferometry configurations. MGD involves straightforward computation per iteration which is very simple to implement compared to alternating minimization schemes. As illustrated in our work MGD showed excellent object wave recoveries when interferograms in different configurations (on axis or off axis) were used. For the step phase object used, MGD showed excellent phase step recovery indicating full pixel resolution performance. Since MGD effectively utilized the expected object wave sparsity for phase reconstruction, the rms phase error performance better than the usual definition of single-pixel based shot-noise level was observed. A detailed study of this aspect will be carried out in future. We believe that a robust approach MGD can lead to widespread employment of optimization based methodologies in interferometry and digital holographic imaging applications. The associated devices can thus operate in single-shot mode with full pixel resolution performance as well as superior accuracy. MGD as a concept can potentially work for a number of optimization problems as we will explore in future.

\end{document}